\documentclass[%
 reprint, 
 amsmath,amssymb,
 aps,
]{revtex4-2}
\usepackage{graphicx}
\usepackage{dcolumn}
\usepackage{bm}
\usepackage{xcolor}

\usepackage{fancyhdr}
\fancypagestyle{firstpage}{
  \fancyhf{}
  \fancyhead[R]{\small IPMU26-0024}
  
}

\begin{document}
\title{Wick-connected theories and Lorentz violation } 

\author{Niklas G\"{o}tzberger}
\affiliation{Arnold Sommerfeld Center for Theoretical Physics,
Ludwig–Maximilians–Universität München,
Theresienstraße 37, 80333 Munich, Germany.}

\author{Anamaria Hell}
\affiliation{
 Kavli IPMU (WPI), UTIAS, The University of Tokyo,
and Center for Data-Driven Discovery,
5-1-5 Kashiwanoha, Kashiwa, Chiba 277-8583, Japan 
}%

\begin{abstract}

We consider double Wick rotation in field theories, which analytically continues the time coordinate, and then reinterprets one of the spatial directions as the new Lorentzian time. We show that if Lorentz-invariance is absent, Wick-connected theories are no longer necessarily equivalent. Focusing on flat spacetime, we study the propagating modes, unitarity and renormalizability of such Wick-connected theories, and provide criteria for when such notions fail to translate. 
\end{abstract}

\maketitle
\thispagestyle{firstpage}

\section{Introduction}

Wick rotation is one of the most widely used tools, with applications in theoretical physics ranging from quantum fields, string theory, gravity to cosmology and statistical physics. It is most commonly used as a mathematical trick that relates spacetimes with Lorentzian and Euclidean signature by analytically continuing the time coordinate \cite{Peskin:1995ev}. 

When performing this procedure, it is natural to question how to relate the theories written in the Euclidean space to their Lorentzian partners, obtained after the Wick rotation. If a theory is defined in the spacetime with Lorentzian signature, time is defined relative to the other coordinates as a coordinate with an opposite sign. In contrast, in the Euclidean space, the same notion is absent, as all coordinates appear with an equal relative signature. Therefore, it is interesting to explore whether theories obtained by Wick rotating different Euclidean coordinates will yield physically indistinguishable results. In this work, we will explore this question, and referring to the theories that are connected through these rotations as \textit{Wick-connected theories}. 

Being an integral part of many derivations, Wick rotation was been extensively studied over the years. For Lorentz-invariant quantum field theories, starting with the works of Dyson and Schwinger \cite{Dyson:1949ha, Schwinger:1958qau, Schwinger:1959zz}, its rigorous formulation was given by the  Osterwalder-Schrader reconstruction theorem \cite{Osterwalder:1973dx, Osterwalder:1974tc}, while the conditions for its existence, associated challenges, and possible extensions when generalizing to the gravitational theories were further studied in \cite{Helleland:2015wva, Samuel:2015oea, Visser:2017atf, Baldazzi:2018mtl, Baldazzi:2019kim, Kontsevich:2021dmb, Witten:2021nzp, Visser:2021ucg}.  In the context of performing a Wick rotation from Lorentzian- to Euclidean-signature spaces, and then further extending the theory to new spacetime with the Lorentzian signature, \textit{double Wick rotation} (DWR) has been particularly interesting. This trick has mostly been applied in string theory \cite{Arutyunov:2007tc,Arutyunov:2009zu,Arutyunov:2014cra,Arutyunov:2014jfa} and general relativity \cite{Jones:2004rg,Astorino:2026kuv,Dai:2025yhi, Colleaux:2025uiw}, where the two Wick rotations are performed such that time and space coordinates on the world-sheet or spacetime exchange the roles. Curiously, as pointed out in \cite{Arutyunov:2007tc}, if the theories are not Lorentz-invariant, the mirror Hamiltonian obtained by this trick is not the same as the original one. However, one could still expect that the dispersion relation and S-matrix for the mirror theory can be obtained after DWR of the original ones. Moreover, DWR can relate various black-hole and cosmological solutions, mapping spacetime with one symmetry into another with a different symmetry \cite{Colleaux:2025uiw}. 

Nevertheless, one could naturally expect that at least in some cases, the properties of a given theory such as its number of degrees of freedom, unitarity or renormalizability would change upon DWR, giving rise to new physics. However, to our knowledge, a study of how and whether these properties translate was never done. In this work, we will fill this gap, and explore this question by considering a series of examples of Lorentz invariant, and Lorentz violating theories, with the aim of classifying when one can expect the translation of properties between Wick-connected theories to break down. In particular, we will show that those that violate Lorentz-invariance are especially interesting, and demonstrate how in such cases the fundamental properties change.

\textit{Conventions:}  We will use Greek indices $\mu,\nu=0,1,2,3$ for spacetime coordinates with Lorentz signature, big Latin letters $A,B=1,2,3,4$ for Euclidean signature.

\section{The basic idea}
Time plays a key role in defining physical theories. While in spacetime with Lorentz signature it is determined through a relative sign, in Euclidean space such notion is lost. Intrigued by this, in this section we will consider two theories of scalar fields and check whether their properties would change depending on which coordinate is Wick-rotated.

As a first step, let us consider the theory of a scalar field, defined in the Euclidean space: 
\begin{equation}
    S=\frac{i}{2}\int d^4x_E\left(\partial_{A}\phi\partial^{A}\phi\right)=iS_E,
\end{equation}
where $\partial_{A}$, is the partial derivative with respect to the coordinate $x^A$, $S_E$ denotes the standard Euclidean action, and the background is given by:  
\begin{equation}
    ds^2=\delta_{AB}dx^Adx^B,\qquad A,B=1,2,3,4
\end{equation}
We can notice that if we identify the $x_4$ coordinate as the time component, and the remaining ones as spatial by setting $x_4=-it$ and $x_{A/\{4\}}=x_i$, $i=1,2,3$, we find: 
\begin{equation}
    S=-\frac{1}{2}\int d^4x\left(\partial_{\mu}\phi\partial^{\mu}\phi\right)=\frac{1}{2}\int d^4x\left(\dot{\phi}^2-\partial_i\phi\partial_i\phi\right),
\end{equation}
where dot denotes the derivative with respect to the coordinate time $t$. However, if we have instead identified any other coordinate among $\{x_A\}$ as time while performing the Wick rotation ( for example, $x_3=-it$ instead of the $x_4$ coordinate), then we would arrive at exactly the same result. In other words, the above theory is \textit{invariant under the choice of time upon Wick rotation}. One can easily show that the same analysis can also hold for more complicated theories, such as Maxwell's electrodynamics, provided that they have an O(4) symmetry in the Euclidean space. Let us now consider the case when this invariance is explicitly broken.

Lorentz-violation is an curious possibility that may realize in nature \cite{Kostelecky:2008ts, Mattingly:2005re, PierreAuger:2026dip}, providing one of the directions for theories beyond the standard model of particle physics \cite{Kostelecky:2003fs,Colladay:1998fq,Seifert:2018mmr,Junior:2026gta} and Einstein's gravity \cite{Kostelecky:2003fs,Alfaia:2026swy,Kostelecky:2017zob,Bellorin:2013zbp} on scales that we have not yet probed. Among such theories, Ho\v{r}ava gravity is especially intriguing, as it provides a renormalizable theory of gravity \cite{Horava:2009if,Horava:2009uw, Barvinsky:2015kil,Anselmi:2007ri,Blas:2009qj,Blas:2010hb} and interesting cosmology \cite{Calcagni:2009ar,Kiritsis:2009sh,Vilenkin:1982de,Pereira:2007yy, Mukohyama:2010xz}. Such violation also naturally appears in theories with non-minimal coupling to gravity, that are cosmologically interesting as possible dark energy candidates and modify the speed of propagation of the gravitational waves \cite{Golovnev:2008hv,Kobayashi:2009hj,DeFelice:2025ykh, DeFelice:2025khe, Hell:2026sxt}.

Let us now explore if the properties of Wick connected theories whose Lorentz invariance is explicitly broken change upon selecting different coordinates as time. In Euclidean space, this corresponds to breaking the O(4) rotational group. As an example where this takes place, let us consider the following scalar field theory in Euclidean space:
  \begin{equation}\label{scf4}
        S=iS_E=-\frac{i}{2}\int d^4x_E\left[\partial_{A}\phi\partial^{A}\phi-\alpha \partial_2\phi\partial_2\phi\right],
\end{equation}
where $\alpha$ is a parameter. By choosing $x_4=-it$, we find: 
\begin{equation}\label{LIVscalar1}
    S=-\frac{1}{2}\int d^4x\left[\partial_{\mu}\phi\partial^{\mu}\phi-\alpha \partial_2\phi\partial_2\phi\right].
\end{equation}
We would also obtain the same result if we would instead choose $x_3$ or $x_1$ as a time coordinate after Wick rotation. However, if we instead set $ x_2=-ix_0, $ 
and accordingly keep the remaining coordinates $x_i$ as spatial ones with $i=1,3,4$, we find: 
\begin{equation}
  \begin{split}
        S=-\frac{1}{2}\int d^4x\left[-(1-\alpha) \partial_2\phi\partial_2\phi+\partial_i\phi\partial_i\phi\right].
  \end{split}
\end{equation}
For $\alpha=1,$ the scalar field in the above action loses its kinetic term and stops to propagate, leading to the change in the number of degrees of freedom (\textit{dof}): the Wick-connected action (\ref{LIVscalar1}) with  $\alpha=1$ describes one propagating mode with dispersion relation $\omega^2=k_1^2+k_3^2$. The above example of scalar field is therefore \textit{not invariant under the choice of time under Wick rotation. } By changing the time-coordinate, their defining properties can change.

For $\alpha\neq1$, rescaling of $x_2$ coordinates restores Lorentz invariance if the theory is free.  Otherwise, in the presence of coupling with other fields Lorentz-violation would translate there. Therefore, even if the number of \textit{dof} would remain the same, the properties of theories can change. In the following, we will analyze how this occurs through several examples.

\section{When ghosts emerge}

Previously we have seen that Wick-connected theories with no Lorentz invariance can describe different number of \textit{dof}. However, this can be even more drastic if the dispersion relation of the fields is contains higher momentum-dependent terms. Such contributions appear in Ho\v{r}ava-Lifshitz gravity or minimally modified theories of gravity \cite{Lin:2017oow}. Here, we will illustrate the effects of such terms on the following example in the Euclidean space: 
\begin{equation}
    S=iS_E=-\frac{i}{2}\int d^4x_E\left[\partial_{A}\phi\partial^{A}\phi+\partial_i\partial_i\phi\partial_j\partial_j\phi\right]. 
\end{equation}
If we choose to Wick-rotate the $x_4$ coordinate, we will find a scalar field with higher-order momentum terms: 
\begin{equation}
    S=-\frac{1}{2}\int d^4x\left[\partial_{\mu}\phi\partial^{\mu}\phi+\partial_i\partial_i\phi\partial_j\partial_j\phi\right]
\end{equation}
However, if we instead choose any of the other coordinates, we can easily notice that the theory will completely change.  For example, by setting $x_1=-it$, we find:
\begin{equation}
    \begin{split}
        S&=-\frac{1}{2}\int d^4x\Big[\Ddot{\phi}\Ddot{\phi}+2\Ddot{\phi}(\partial_2^2+\partial_3^2)\phi\\&+(\partial_2^2+\partial_3^2)\phi(\partial_2^2+\partial_3^2)\phi+\partial_{\mu}\phi\partial^{\mu}\phi\ \Big], 
    \end{split}
\end{equation}
where dot stands for the derivative with respect to time. The above theory describes an Ostrogradsky ghost \cite{Ostrogradsky:1850fid}, together with a healthy scalar mode. This implies that unitarity can be easily violated if the Lorentz violation is broken in such a way that higher-order derivatives along a preffered direction appear.

\section{Strong coupling }\label{Strong coupling}

Another very interesting case appears if we consider the following action \cite{Dvali:2005nt}:
\begin{equation}\label{vectorLI}
    S=\int d^4x\left(-\frac{1}{4}F_{\mu\nu}F^{\mu\nu}-\frac{1}{2}m^2A_iA_i\right). 
\end{equation}
This theory propagates two \textit{dof}, with derivation given in the   Appendix \ref{Appendix B}:
\begin{equation}
    \begin{split}
         \mathcal{L}_{LvP_1}&=\frac{1}{2}\left[\dot{A}^T_i\dot{A}^T_i-\partial_iA_j^T\partial_iA_j^T-m^2A_i^TA_i^T\right]. 
    \end{split}
\end{equation}
By Wick rotating  $ t=ix_4,$ and $A_0=-iA_4$, we arrive at: 
\begin{equation}
    S=iS_E=i\int d^4x_E-\frac{1}{4}F_{AB}^EF^{EAB}-\frac{1}{2}m^2A_iA_i. 
\end{equation}
Next we perform another rotation, setting $x_1$ coordinate as the new time coordinate through $x_1=-it$ and $A_1=iA_0$. 
In this case, the action becomes:
\begin{equation}\label{DRWvec}
    \mathcal{L}_{LvP_2}=-\frac{1}{4}F_{\mu\nu}F^{\mu\nu}+\frac{1}{2}m^2\left(A_0A_0-A_2A_2-A_3A_3\right), 
\end{equation}
where $\mu=0,2,3,4$. As a result, the structure of the constraints change as $A_0$ component now acquires a mass term. Moreover, the system is no longer spherically symmetric, but rather has a spatial anisotropy in the $x_4$ direction, and an isotropic subspace defined by $x_2-x_3$ plane. To find the propagating modes, we decompose the fields into even and odd modes $A_4=\partial_4A$, $A_a=\partial_a\chi+\varepsilon_{ab}\partial_b v$ which decouple.  Following then the standard recipe for counting the \textit{dof} \cite{Hell:2026blj} with steps described in detail in the Appendix \ref{Appendix B} we find in the momentum space:
\begin{equation}
    \tilde{\mathcal{L}}_{odd}=\frac{(k_2^2+k_3^2)}{2}\left(\dot{v}_{\vec{k}}\dot{v}_{\vec{-k}}-(\vec{k}^2+m^2)v_{\vec{k}} v_{\vec{-k}}\right), 
\end{equation}
and two even ones after the diagonalization: 
\begin{widetext}
\begin{equation}\label{Even LIV Proca Momentum space}
\begin{split}
\tilde{\mathcal{L}}_{\text{even}} &= 
\frac{k_4^{2} m^{2}}{2 k_4^{2}+2 m^{2}} \dot{A}_{\vec{k}}\dot{A}_{-\vec{k}} + \frac{(k_2^{2}+k_3^{2})(k_4^{2}+m^{2})}{2 k_2^{2}+2 k_3^{2}+2 k_4^{2}+2 m^{2}}
\dot{\chi}_{2\vec{k}}\dot{\chi}_{2(-\vec{k})} 
 + \frac{m^{2} k_4^{2} (k_2^{2}+k_3^{2})}{k_4^{2}+m^{2}}
(\chi_{2\vec{k}} A_{-\vec{k}}+\chi_{2(-\vec{k})} A_{\vec{k}}) \\
& - \frac{(k_2^{2}+k_3^{2})(m-k_4)(k_4+m) k_4^{2} m^{2}}
{2 (k_4^{2}+m^{2})^{2}} A_{\vec{k}} A_{-\vec{k}} 
 + \frac{(k_2^{2}+k_3^{2})(m-k_4)(k_4+m)}{2}
\chi_{\vec{k}}\chi_{-\vec{k}}
\end{split}
\end{equation}
\end{widetext}
In the above, $\vec{k}^2=k_2^2+k_3^2+k_4^2$. We can notice that the scalar $A$ comes multiplied with the mass term, and it dissapears when $m=0$. As a result, on very high energies where the Lorentz invariance is broken only slightly one could expect its minimal amplitude of quantum fluctuations to be singular in mass. This is commonly the case in massive gauge theories, where new modes appear upon modifying them with a mass term \cite{Vainshtein:1972sx, Gruzinov:2001hp, Deffayet:2001uk, Chamseddine:2018gqh,Hell:2021wzm,  Hell:2021oea, Hell:2025pso}. Following this analogy, in the presence of self-interacting terms such as $A_{\mu}A^{\mu}\partial^{\nu}A_{\nu}$ or $(A_{\mu}A^{\mu})^2$, once the non-linear terms become of the same order as linear ones, one could therefore expect that this mode will become strongly coupled, and decouple from the remaining two modes up to small corrections. Such mechanism is similar to the Vainshtein mechanism in massive gravity \cite{Vainshtein:1972sx}, which was also extended to massive mimetic gravity as well as massive vector and two-form fields. Through a quick estimate, and assuming that the coupling $\lambda$ is smaller than unity, one can show that for $\lambda A_{\mu}A^{\mu}\partial^{\nu}A_{\nu}$ one could expect this strong coupling length-scale to be $L_{str}\sim \lambda^{-\frac{1}{3}}m^{-1}$ for $k_2^2\sim k_3^2\sim k_4^2\sim \frac{1}{L^2}\gg m^2$.

\section{Renormalization}

Let us now investigate whether renormalizability can be translated between different Wick-connected theories. As a first case, let's consider in the Euclidean space: 
\begin{equation}
    S_E=\int d^4x\left(\partial_A\phi\partial^A\phi+\frac{1}{2M^2}\partial_1^2\phi\partial_1^2\phi-\frac{m^2}{2}\phi^2-\frac{\lambda}{4}\phi^4\right)
\end{equation}
One can  check that the Wick-connected theories obtained by choosing any of the coordinates as time are power-counting renormalizable. Thus, in this case, renormalizability is translated. However, if one sets $x_1$ coordinate as time through $x_1=-it$, then one will obtain a ghost mode in addition to a healthy scalar. 

Let us also see what would happen to the vector field theory (\ref{vectorLI}), coupled to the conserved source $\mathcal{L}_{int}=-A_{\mu}J^{\mu}$. Written in the original form, it is renormalizable. However, if one performs the DWR such that $t=ix_4$, and then $x_1=-i\tilde{t}$, one finds (\ref{DRWvec}), which propagates an additional mode. This mode is very similar to the longitudinal mode of the Proca theory as it disappears when the mass is set to zero. In the case of Proca fields, the theory is renormalizable if the source is conserved, as then the part of the propagator that tends to a constant at high energies cancels out \cite{PhysRev.126.1563}. 

Let us show that the same condition takes place here as well. For this, it is convenient to write (\ref{DRWvec}) as: 
\begin{equation}\label{DRWvec2}
    \mathcal{L}_{LvP_2}=\frac{1}{2} A^{\mu}(\eta_{\mu\nu}(\partial_{\gamma}\partial^{\gamma}-m^2)-\partial_{\mu}\partial_{\nu}+m^2n_{\mu}n_{\nu})A^{\nu},
\end{equation}
where $n_{\mu}=\delta_{\mu}^4$. The propagator for this theory reads
\begin{equation}\label{propagatorVec}
\begin{split}
    D^{\mu\nu}=\frac{i}{k^2+m^2}\Big[-\eta^{\mu\nu}+\frac{k^\mu k^\nu}{m^2}\frac{k^2}{\tilde\Delta}\\+\frac{m^2n^\mu n^\nu}{\tilde\Delta}+\frac{n_\rho k^\rho}{\tilde \Delta}(k^\mu n^\nu+n^\mu k^\nu)\Big]
\end{split}
\end{equation}
where $\tilde\Delta\equiv -k^2+(n_\mu k^\mu)^2$. One can easily see that this expression reduces to the ordinary Proca propagator for one sets $n= 0$. Moreover, we can notice that the same condition as in the previous case holds: if the source is conserved, the second term in (\ref{propagatorVec}), which becomes constant at high energies and thus indicates power-counting non-renormalizability, drops out. The standard vector propagator and $m^2n^\mu n^\nu$ term survive.

\section{Compactifications} 

It is also interesting to note that the Wick-connected theories can yield different results when Lorenz-symmetry is spontaneously broken. As an example of this effect, we can consider a theory of canonical scalar fields in five dimensions in which the $x_5$ coordinate is compactified $S_E=-\frac{1}{2}\int d^5x_E\partial_A\phi\partial^A\phi$. If we set any coordinate apart $x_5$ to be time upon Wick rotation, this can be seen as spontaneous Lorentz-symmetry breaking, with an the action that describes a tower of massive scalar fields once one decomposes the scalar in Fourier modes and integrates over $x_5$. However, if one instead selects $x_5$ as time upon the Wick rotation, the theories no longer describe the same physics. In particular, the resulting theory has compact time, and thus closed time-like curves, which leads to the violation of causality.

\section{Implications and Discussion} 
In this work, we have considered a series of examples formulated in flat spacetime, and observed how properties of Wick-connected theories change under the double Wick rotation. Based on them, we provide the following guidelines on what to expect when performing the DWR: 

\textit{1. Degrees of freedom and symmetries:} One of the immediate changes when choosing different coordinates as time is the change in the number of propagating modes. 
The criterion to examine is the symmetry of the theory: If the theory has O(4) invariance in Euclidean space, or, analogously is Lorentz-invariant in the corresponding Wick-connected version, then the theory is invariant under the choice of time, meaning that its properties are translated. However, if instead the theory breaks this symmetry, then not all time choices are equivalent and only the subset of Wick-connected theories related to the residual symmetry have their properties translated. 
For example, the scalar field theory in (\ref{scf4}) with $\alpha=1$ will describe a scalar mode if one Wick rotates in $x_1$, $x_3$ or $x_4$ due to the O(3) invariance in the Euclidean space. The resulting Wick-connected theories would describe a scalar mode, and break isotropy, having only an isotropic subspace. However, if one instead chooses $x_2$ as time, the resulting theory has invariance under the group of rotation, but it has no propagating modes.

\textit{2. Emergence of ghosts} As we have seen, the appearance of extra \textit{dof} may also result in the appearance of Ostrogradsky modes in the theory. Whenever the theory has higher-order spatial derivatives, one can expect that at least one Wick-connected theory contains modes appearing with higher-order derivatives in time.  While we have demonstrated this on a simple example of a scalar field, one can also expect it to emerge in more complicated theories such as theories that are Wick-connected to Ho\v{r}ava gravity \cite{Horava:2009if,Horava:2009uw}. It is being built on foliation preserving diffeomorphisms, with action that contains a potential being built from the spatial metric, and allowing for arbitrarily high derivatives while not introducing any time-derivatives on the metric. While in this formulation, the notion of time is special, one could nevertheless perform the Wick rotation as in \cite{Barvinsky:2015kil}. However, once in the Euclidean space, one can show that another possible Wick rotation which results in a real action is 
\begin{align}
    x_1\rightarrow -it,\qquad N_1^{(E)}\rightarrow +iN_1^{(L)},\qquad \gamma_{1a}\rightarrow i\gamma_{1a}, 
\end{align}
which changes the symmetry and nature of the propagating modes. 

\textit{3. Strong coupling:} Similarly to massive gauge theories, if one of the Wick-connected theories that breaks the Lorentz invariance has more modes than the Lorentz-invariant version which is obtained by setting the parameter characterizing the Lorentz-invariance to zero, it may happen that such modes will be strongly coupled and decouple from the remaining modes in the case of self-interactions. We have demonstrated one case where this might take place, however, more generally, it is related to modifying the constraint structure upon DWR that causes the emergence of extra modes in the first place. However, if the same would become strongly coupled, it would be very interesting to show if the theory beyond the scale analogous to the Vainshtein radius then matches with the Wick-connected partner that propagates less modes. 

\textit{4. Renormalization:} While the structure of theories entirely changes, this does not imply that the renormalizability property is guaranteed to not be translated. If we find a symmetry or conservation law to translate, so do we expect renormalizability to. In our case, source conservation implies that both Wick--connected vector theories will be powercounting--renormalizable.

Overall, DWR has been used in many theories, relating different solutions. In this work, we have explored whether the properties among different Wick connected theories might translate, or if the structure changes entirely. We have shown that it is the latter -- not only do the symmetries change, but the theories describe different number of propagating modes, which can even lead to ghost \textit{dof}. It is still nevertheless interesting to note that the properties may be translated if certain conditions are met. While our analysis was restricted to the cases in flat spacetime, it would be especially interesting to generalize this further to modified theories of gravity, which are cosmologically well-motivated, but can naturally lead to Lorentz violation in some formulations such as with non-minimal coupling to gravity, or explore relating $CP$- and $CT$-invariant theories under DWR that might be intriguing for particle physics phenomenology.

\textit{Acknowledgements}
The authors would like to thank Gia Dvali for pointing out that compactifications with DWR also give rise to inequivalent physics.  AH would also like to thank to Robert Brandenberger, Elisa G. M. Ferreira, Dieter L\"{u}st,  Viatcheslav Mukhanov, Shinji Mukohyama, Taizan Watari, and Tsutomu Yanagida for useful discussions and correspondences. 
AH is supported by JSPS KAKENHI Grant No. JP26K17133, and by the CD3 Google Seed grant.

\appendix

\section{Degrees of freedom of LV Proca theory}\label{Appendix B}
In the following, we will demonstrate the \textit{dof} for the Lorentz-violating Proca theory, given in (\ref{vectorLI}), and after that also how the \textit{dof} after performing the DWR given in (\ref{DRWvec}). 

Let us first consider (\ref{vectorLI}). Due to the spherical symmetry, to find out its \textit{dof}, we can first decompose the spatial part of the vector field into the longitudinal and transverse modes:
\begin{equation}
A_i=A_i^T+\partial_i\chi,\qquad\text{where}\qquad \partial_iA_{i}^T=0. 
\end{equation}
Then, the Lagrangian density becomes:
\begin{equation}
    \begin{split}
         \mathcal{L}_{LvP_1}&=\frac{1}{2}\left[-A_0\Delta A_0+2A_0\Delta\dot{\chi}\right]\\
         &+\frac{1}{2}\left[\dot{A}^T_i\dot{A}^T_i-\partial_iA_j^T\partial_iA_j^T-m^2A_i^TA_i^T\right]
         \\
         &+\frac{1}{2}\left[\partial_i\dot{\chi}\partial_i\dot{\chi}-m^2\partial_i\chi\partial_i\chi\right]
    \end{split}
\end{equation}
where the dot now denotes the derivative with respect to the coordinate time $t$. We can notice that similarly to electrodynamics, $A_0$ is a non-propagating field, which satisfies the constraint:
\begin{equation}
    -\Delta A_0=-\Delta\dot{\chi},
\end{equation}
whose solution is given by:
\begin{equation}
    A_0=\dot{\chi}
\end{equation}
By substituting this solution back to the action, we find:
\begin{equation}
    \begin{split}
         \mathcal{L}_{LvP_1}&=\frac{1}{2}\left[\dot{A}^T_i\dot{A}^T_i-\partial_iA_j^T\partial_iA_j^T-m^2A_i^TA_i^T\right]\\
         &
         -\frac{1}{2}m^2\partial_i\chi\partial_i\chi
    \end{split}
\end{equation}
Therefore, we find that $\chi$ loses its kinetic term and becomes a constrained field, which satisfies:
\begin{equation}
    \Delta\chi=0. 
\end{equation}
Hence, by substituting this back to the Lagrangian density, we find that only two \textit{dof} propagate, corresponding to the two polarizations of the electromagnetic wave, which in contrast to the standard electrodynamics has acquired a mass due to the Lorentz-violating term:
\begin{equation}
    \begin{split}
         \mathcal{L}_{LvP_1}&=\frac{1}{2}\left[\dot{A}^T_i\dot{A}^T_i-\partial_iA_j^T\partial_iA_j^T-m^2A_i^TA_i^T\right]. 
    \end{split}
\end{equation}

Let us now compare this with (\ref{DRWvec}). 
We can notice that after DWR the structure of the constraints change, since the $A_0$ component now acquires a mass term. Moreover, the system is no longer spherically symmetric. While the theory has spatial anisotropy in the $x_4$ direction, it is still isotropic in the subspace defined by $x_2-x_3$. Therefore, by respecting the symmetries of the theory, it could be best to decompose the vector field according to the even and odd modes, that are commonly used in the Bianchi Type I spacetime:
\begin{equation}
    A_4=\partial_4A\qquad A_a=\partial_a\chi+\varepsilon_{ab}\partial_b v,
\end{equation}
where $a,b=2,3$, and $\varepsilon_{ab}$ is the 2-dimensional Levi-Civita symbol. In the above, $A$ and $\chi$ are the even modes, and $v$ is the odd mode. By substituting this decomposition into the above action, and by varying it with respect to $A_0$, we find that $A_0$ satisfies the following constraint:
\begin{equation}
    (-\Delta+m^2)A_0=-\left(\partial_4^2\dot{A}+\Tilde{\Delta}\dot{\chi}\right),
\end{equation}

where $\Delta=\partial_2^2+\partial_3^2+\partial_4^2$, and $\Tilde{\Delta}=\partial_2^2+\partial_3^2$. By solving the constraint, and substituting back to the action we find the following Lagrangian density:
\begin{equation}
    \mathcal{L}_{LvP_2}=\mathcal{L}_{even}+\mathcal{L}_{odd}
\end{equation}
where
\begin{equation}\label{Even LIV Proca}
\begin{split}
\mathcal{L}_{\text{even}} = \frac{1}{2} \Big[
& - \big( \partial_4^2 \dot{A} + \tilde{\Delta} \dot{\chi} \big)
  \frac{1}{-\Delta + m^2}
  \big( \partial_4^2 \dot{A} + \tilde{\Delta} \dot{\chi} \big) \\
& - \dot{A}\, \partial_4^2 \dot{A}
  - \dot{\chi}\, \tilde{\Delta} \dot{\chi}  + A\, \partial_4^4 A
  - A\, \partial_4^2 \Delta A\\
  &
  + 2 A\, \partial_4^2 \tilde{\Delta} \chi  + \chi\, \tilde{\Delta}^2 \chi
  - \chi\, \Delta \tilde{\Delta} \chi
  + m^2 \chi\, \tilde{\Delta} \chi
\Big]
\end{split}
\end{equation}
and 
\begin{equation}
    \mathcal{L}_{odd}=-\frac{1}{2}\left(\dot{v}\Tilde{\Delta}\dot{v}-v\Tilde{\Delta}\Delta v+m^2v\Tilde{\Delta}v\right). 
\end{equation}

Therefore, we can see that the even and odd modes decouple, and that the theory now describes three \textit{dof} -- two even modes and an odd one. The odd mode can be further canonically normalized:
\begin{equation}
    v_n=\sqrt{-\Tilde{\Delta}}v
\end{equation}
after which we find a canonical mode: 
\begin{equation}\label{Odd LIV Proca}
    \mathcal{L}_{odd}=\frac{1}{2}\left(\dot{v}_n^2+v_n\Delta v_n-m^2v_n^2\right). 
\end{equation}
To infer the nature of the even modes, it would be better to diagonalize the kinetic matrix between $A$ and $\chi$. This is possible to achieve by expanding the fields in the Fourier modes, 
\begin{equation}
    X=\int\frac{d^3k}{(2\pi)^{3/2}}e^{i(k_2x_2+k_3x_3+k_4x_4)}X_{\Vec{k}}
\end{equation}
and substituting
\begin{equation}
    \chi_{\Vec{k}}=\chi_{2\Vec{k}}+\frac{k_4^2 }{k_4^2+m^{2}}A_{\Vec{k}}. 
\end{equation}
The Lagrangian density in the momentum space then becomes: 
\begin{widetext}
\begin{equation}\label{Even LIV Proca Momentum space}
\begin{split}
\mathcal{L}_{\text{even}} &= 
\frac{k_4^{2} m^{2}}{2 k_4^{2}+2 m^{2}} \dot{A}_{\vec{k}}\dot{A}_{-\vec{k}} + \frac{(k_2^{2}+k_3^{2})(k_4^{2}+m^{2})}{2 k_2^{2}+2 k_3^{2}+2 k_4^{2}+2 m^{2}}
\dot{\chi}_{2\vec{k}}\dot{\chi}_{2(-\vec{k})} 
 + \frac{m^{2} k_4^{2} (k_2^{2}+k_3^{2})}{k_4^{2}+m^{2}}
(\chi_{2\vec{k}} A_{-\vec{k}}+\chi_{2(-\vec{k})} A_{\vec{k}}) \\
 &- \frac{(k_2^{2}+k_3^{2})(m-k_4)(k_4+m) k_4^{2} m^{2}}
{2 (k_4^{2}+m^{2})^{2}} A_{\vec{k}} A_{-\vec{k}} 
+ \frac{(k_2^{2}+k_3^{2})(m-k_4)(k_4+m)}{2}
\chi_{\vec{k}}\chi_{-\vec{k}}
\end{split}
\end{equation}
\end{widetext}
Therefore, we can see that the theory propagates an entirely different number of \textit{dof}, and has different symmetries then the starting SO(3) invariant theory.

\bibliography{bib.bib}

\begin{thebibliography}{60}%
\makeatletter
\providecommand \@ifxundefined [1]{%
 \@ifx{#1\undefined}
}%
\providecommand \@ifnum [1]{%
 \ifnum #1\expandafter \@firstoftwo
 \else \expandafter \@secondoftwo
 \fi
}%
\providecommand \@ifx [1]{%
 \ifx #1\expandafter \@firstoftwo
 \else \expandafter \@secondoftwo
 \fi
}%
\providecommand \natexlab [1]{#1}%
\providecommand \enquote  [1]{``#1''}%
\providecommand \bibnamefont  [1]{#1}%
\providecommand \bibfnamefont [1]{#1}%
\providecommand \citenamefont [1]{#1}%
\providecommand \href@noop [0]{\@secondoftwo}%
\providecommand \href [0]{\begingroup \@sanitize@url \@href}%
\providecommand \@href[1]{\@@startlink{#1}\@@href}%
\providecommand \@@href[1]{\endgroup#1\@@endlink}%
\providecommand \@sanitize@url [0]{\catcode `\\12\catcode `\$12\catcode `\&12\catcode `\#12\catcode `\^12\catcode `\_12\catcode `\%12\relax}%
\providecommand \@@startlink[1]{}%
\providecommand \@@endlink[0]{}%
\providecommand \url  [0]{\begingroup\@sanitize@url \@url }%
\providecommand \@url [1]{\endgroup\@href {#1}{\urlprefix }}%
\providecommand \urlprefix  [0]{URL }%
\providecommand \Eprint [0]{\href }%
\providecommand \doibase [0]{https://doi.org/}%
\providecommand \selectlanguage [0]{\@gobble}%
\providecommand \bibinfo  [0]{\@secondoftwo}%
\providecommand \bibfield  [0]{\@secondoftwo}%
\providecommand \translation [1]{[#1]}%
\providecommand \BibitemOpen [0]{}%
\providecommand \bibitemStop [0]{}%
\providecommand \bibitemNoStop [0]{.\EOS\space}%
\providecommand \EOS [0]{\spacefactor3000\relax}%
\providecommand \BibitemShut  [1]{\csname bibitem#1\endcsname}%
\let\auto@bib@innerbib\@empty
\bibitem [{\citenamefont {Peskin}\ and\ \citenamefont {Schroeder}(1995)}]{Peskin:1995ev}%
  \BibitemOpen
  \bibfield  {author} {\bibinfo {author} {\bibfnamefont {M.~E.}\ \bibnamefont {Peskin}}\ and\ \bibinfo {author} {\bibfnamefont {D.~V.}\ \bibnamefont {Schroeder}},\ }\href {https://doi.org/10.1201/9780429503559} {\emph {\bibinfo {title} {{An Introduction to quantum field theory}}}}\ (\bibinfo  {publisher} {Addison-Wesley},\ \bibinfo {address} {Reading, USA},\ \bibinfo {year} {1995})\BibitemShut {NoStop}%
\bibitem [{\citenamefont {Dyson}(1949)}]{Dyson:1949ha}%
  \BibitemOpen
  \bibfield  {author} {\bibinfo {author} {\bibfnamefont {F.~J.}\ \bibnamefont {Dyson}},\ }\bibfield  {title} {\bibinfo {title} {{The S matrix in quantum electrodynamics}},\ }\href {https://doi.org/10.1103/PhysRev.75.1736} {\bibfield  {journal} {\bibinfo  {journal} {Phys. Rev.}\ }\textbf {\bibinfo {volume} {75}},\ \bibinfo {pages} {1736} (\bibinfo {year} {1949})}\BibitemShut {NoStop}%
\bibitem [{\citenamefont {Schwinger}(1958)}]{Schwinger:1958qau}%
  \BibitemOpen
  \bibfield  {author} {\bibinfo {author} {\bibfnamefont {J.}~\bibnamefont {Schwinger}},\ }\bibfield  {title} {\bibinfo {title} {{ON THE EUCLIDEAN STRUCTURE OF RELATIVISTIC FIELD THEORY}},\ }\href {https://doi.org/10.1073/pnas.44.9.956} {\bibfield  {journal} {\bibinfo  {journal} {Proc. Nat. Acad. Sci.}\ }\textbf {\bibinfo {volume} {44}},\ \bibinfo {pages} {956} (\bibinfo {year} {1958})}\BibitemShut {NoStop}%
\bibitem [{\citenamefont {Schwinger}(1959)}]{Schwinger:1959zz}%
  \BibitemOpen
  \bibfield  {author} {\bibinfo {author} {\bibfnamefont {J.}~\bibnamefont {Schwinger}},\ }\bibfield  {title} {\bibinfo {title} {{Euclidean Quantum Electrodynamics}},\ }\href {https://doi.org/10.1103/PhysRev.115.721} {\bibfield  {journal} {\bibinfo  {journal} {Phys. Rev.}\ }\textbf {\bibinfo {volume} {115}},\ \bibinfo {pages} {721} (\bibinfo {year} {1959})}\BibitemShut {NoStop}%
\bibitem [{\citenamefont {Osterwalder}\ and\ \citenamefont {Schrader}(1973)}]{Osterwalder:1973dx}%
  \BibitemOpen
  \bibfield  {author} {\bibinfo {author} {\bibfnamefont {K.}~\bibnamefont {Osterwalder}}\ and\ \bibinfo {author} {\bibfnamefont {R.}~\bibnamefont {Schrader}},\ }\bibfield  {title} {\bibinfo {title} {{AXIOMS FOR EUCLIDEAN GREEN'S FUNCTIONS}},\ }\href {https://doi.org/10.1007/BF01645738} {\bibfield  {journal} {\bibinfo  {journal} {Commun. Math. Phys.}\ }\textbf {\bibinfo {volume} {31}},\ \bibinfo {pages} {83} (\bibinfo {year} {1973})}\BibitemShut {NoStop}%
\bibitem [{\citenamefont {Osterwalder}\ and\ \citenamefont {Schrader}(1975)}]{Osterwalder:1974tc}%
  \BibitemOpen
  \bibfield  {author} {\bibinfo {author} {\bibfnamefont {K.}~\bibnamefont {Osterwalder}}\ and\ \bibinfo {author} {\bibfnamefont {R.}~\bibnamefont {Schrader}},\ }\bibfield  {title} {\bibinfo {title} {{Axioms for Euclidean Green's Functions. 2.}},\ }\href {https://doi.org/10.1007/BF01608978} {\bibfield  {journal} {\bibinfo  {journal} {Commun. Math. Phys.}\ }\textbf {\bibinfo {volume} {42}},\ \bibinfo {pages} {281} (\bibinfo {year} {1975})}\BibitemShut {NoStop}%
\bibitem [{\citenamefont {Helleland}\ and\ \citenamefont {Hervik}(2018)}]{Helleland:2015wva}%
  \BibitemOpen
  \bibfield  {author} {\bibinfo {author} {\bibfnamefont {C.}~\bibnamefont {Helleland}}\ and\ \bibinfo {author} {\bibfnamefont {S.}~\bibnamefont {Hervik}},\ }\bibfield  {title} {\bibinfo {title} {{A Wick-rotatable metric is purely electric}},\ }\href {https://doi.org/10.1016/j.geomphys.2017.09.015} {\bibfield  {journal} {\bibinfo  {journal} {J. Geom. Phys.}\ }\textbf {\bibinfo {volume} {123}},\ \bibinfo {pages} {424} (\bibinfo {year} {2018})},\ \Eprint {https://arxiv.org/abs/1504.01244} {arXiv:1504.01244 [math-ph]} \BibitemShut {NoStop}%
\bibitem [{\citenamefont {Samuel}(2016)}]{Samuel:2015oea}%
  \BibitemOpen
  \bibfield  {author} {\bibinfo {author} {\bibfnamefont {J.}~\bibnamefont {Samuel}},\ }\bibfield  {title} {\bibinfo {title} {{Wick Rotation in the Tangent Space}},\ }\href {https://doi.org/10.1088/0264-9381/33/1/015006} {\bibfield  {journal} {\bibinfo  {journal} {Class. Quant. Grav.}\ }\textbf {\bibinfo {volume} {33}},\ \bibinfo {pages} {015006} (\bibinfo {year} {2016})},\ \Eprint {https://arxiv.org/abs/1510.07365} {arXiv:1510.07365 [gr-qc]} \BibitemShut {NoStop}%
\bibitem [{\citenamefont {Visser}(2017)}]{Visser:2017atf}%
  \BibitemOpen
  \bibfield  {author} {\bibinfo {author} {\bibfnamefont {M.}~\bibnamefont {Visser}},\ }\bibfield  {title} {\bibinfo {title} {{How to Wick rotate generic curved spacetime}},\ }\href@noop {} {\  (\bibinfo {year} {2017})},\ \Eprint {https://arxiv.org/abs/1702.05572} {arXiv:1702.05572 [gr-qc]} \BibitemShut {NoStop}%
\bibitem [{\citenamefont {Baldazzi}\ \emph {et~al.}(2019{\natexlab{a}})\citenamefont {Baldazzi}, \citenamefont {Percacci},\ and\ \citenamefont {Skrinjar}}]{Baldazzi:2018mtl}%
  \BibitemOpen
  \bibfield  {author} {\bibinfo {author} {\bibfnamefont {A.}~\bibnamefont {Baldazzi}}, \bibinfo {author} {\bibfnamefont {R.}~\bibnamefont {Percacci}},\ and\ \bibinfo {author} {\bibfnamefont {V.}~\bibnamefont {Skrinjar}},\ }\bibfield  {title} {\bibinfo {title} {{Wicked metrics}},\ }\href {https://doi.org/10.1088/1361-6382/ab187d} {\bibfield  {journal} {\bibinfo  {journal} {Class. Quant. Grav.}\ }\textbf {\bibinfo {volume} {36}},\ \bibinfo {pages} {105008} (\bibinfo {year} {2019}{\natexlab{a}})},\ \Eprint {https://arxiv.org/abs/1811.03369} {arXiv:1811.03369 [gr-qc]} \BibitemShut {NoStop}%
\bibitem [{\citenamefont {Baldazzi}\ \emph {et~al.}(2019{\natexlab{b}})\citenamefont {Baldazzi}, \citenamefont {Percacci},\ and\ \citenamefont {Skrinjar}}]{Baldazzi:2019kim}%
  \BibitemOpen
  \bibfield  {author} {\bibinfo {author} {\bibfnamefont {A.}~\bibnamefont {Baldazzi}}, \bibinfo {author} {\bibfnamefont {R.}~\bibnamefont {Percacci}},\ and\ \bibinfo {author} {\bibfnamefont {V.}~\bibnamefont {Skrinjar}},\ }\bibfield  {title} {\bibinfo {title} {{Quantum fields without Wick rotation}},\ }\href {https://doi.org/10.3390/sym11030373} {\bibfield  {journal} {\bibinfo  {journal} {Symmetry}\ }\textbf {\bibinfo {volume} {11}},\ \bibinfo {pages} {373} (\bibinfo {year} {2019}{\natexlab{b}})},\ \Eprint {https://arxiv.org/abs/1901.01891} {arXiv:1901.01891 [gr-qc]} \BibitemShut {NoStop}%
\bibitem [{\citenamefont {Kontsevich}\ and\ \citenamefont {Segal}(2021)}]{Kontsevich:2021dmb}%
  \BibitemOpen
  \bibfield  {author} {\bibinfo {author} {\bibfnamefont {M.}~\bibnamefont {Kontsevich}}\ and\ \bibinfo {author} {\bibfnamefont {G.}~\bibnamefont {Segal}},\ }\bibfield  {title} {\bibinfo {title} {{Wick Rotation and the Positivity of Energy in Quantum Field Theory}},\ }\href {https://doi.org/10.1093/qmath/haab027} {\bibfield  {journal} {\bibinfo  {journal} {Quart. J. Math. Oxford Ser.}\ }\textbf {\bibinfo {volume} {72}},\ \bibinfo {pages} {673} (\bibinfo {year} {2021})},\ \Eprint {https://arxiv.org/abs/2105.10161} {arXiv:2105.10161 [hep-th]} \BibitemShut {NoStop}%
\bibitem [{\citenamefont {Witten}(2021)}]{Witten:2021nzp}%
  \BibitemOpen
  \bibfield  {author} {\bibinfo {author} {\bibfnamefont {E.}~\bibnamefont {Witten}},\ }\bibfield  {title} {\bibinfo {title} {{A Note On Complex Spacetime Metrics}},\ }\href@noop {} {\  (\bibinfo {year} {2021})},\ \Eprint {https://arxiv.org/abs/2111.06514} {arXiv:2111.06514 [hep-th]} \BibitemShut {NoStop}%
\bibitem [{\citenamefont {Visser}(2022)}]{Visser:2021ucg}%
  \BibitemOpen
  \bibfield  {author} {\bibinfo {author} {\bibfnamefont {M.}~\bibnamefont {Visser}},\ }\bibfield  {title} {\bibinfo {title} {{Feynman{\textquoteright}s i{\ensuremath{\epsilon}} prescription, almost real spacetimes, and acceptable complex spacetimes}},\ }\href {https://doi.org/10.1007/JHEP08(2022)129} {\bibfield  {journal} {\bibinfo  {journal} {JHEP}\ }\textbf {\bibinfo {volume} {08}},\ \bibinfo {pages} {129}},\ \Eprint {https://arxiv.org/abs/2111.14016} {arXiv:2111.14016 [gr-qc]} \BibitemShut {NoStop}%
\bibitem [{\citenamefont {Arutyunov}\ and\ \citenamefont {Frolov}(2007)}]{Arutyunov:2007tc}%
  \BibitemOpen
  \bibfield  {author} {\bibinfo {author} {\bibfnamefont {G.}~\bibnamefont {Arutyunov}}\ and\ \bibinfo {author} {\bibfnamefont {S.}~\bibnamefont {Frolov}},\ }\bibfield  {title} {\bibinfo {title} {{On String S-matrix, Bound States and TBA}},\ }\href {https://doi.org/10.1088/1126-6708/2007/12/024} {\bibfield  {journal} {\bibinfo  {journal} {JHEP}\ }\textbf {\bibinfo {volume} {12}},\ \bibinfo {pages} {024}},\ \Eprint {https://arxiv.org/abs/0710.1568} {arXiv:0710.1568 [hep-th]} \BibitemShut {NoStop}%
\bibitem [{\citenamefont {Arutyunov}\ and\ \citenamefont {Frolov}(2009)}]{Arutyunov:2009zu}%
  \BibitemOpen
  \bibfield  {author} {\bibinfo {author} {\bibfnamefont {G.}~\bibnamefont {Arutyunov}}\ and\ \bibinfo {author} {\bibfnamefont {S.}~\bibnamefont {Frolov}},\ }\bibfield  {title} {\bibinfo {title} {{String hypothesis for the AdS(5) x S**5 mirror}},\ }\href {https://doi.org/10.1088/1126-6708/2009/03/152} {\bibfield  {journal} {\bibinfo  {journal} {JHEP}\ }\textbf {\bibinfo {volume} {03}},\ \bibinfo {pages} {152}},\ \Eprint {https://arxiv.org/abs/0901.1417} {arXiv:0901.1417 [hep-th]} \BibitemShut {NoStop}%
\bibitem [{\citenamefont {Arutyunov}\ and\ \citenamefont {van Tongeren}(2014)}]{Arutyunov:2014cra}%
  \BibitemOpen
  \bibfield  {author} {\bibinfo {author} {\bibfnamefont {G.}~\bibnamefont {Arutyunov}}\ and\ \bibinfo {author} {\bibfnamefont {S.~J.}\ \bibnamefont {van Tongeren}},\ }\bibfield  {title} {\bibinfo {title} {{$\mathrm{AdS}_5 \times \mathrm{S}^5$ mirror model as a string sigma model}},\ }\href {https://doi.org/10.1103/PhysRevLett.113.261605} {\bibfield  {journal} {\bibinfo  {journal} {Phys. Rev. Lett.}\ }\textbf {\bibinfo {volume} {113}},\ \bibinfo {pages} {261605} (\bibinfo {year} {2014})},\ \Eprint {https://arxiv.org/abs/1406.2304} {arXiv:1406.2304 [hep-th]} \BibitemShut {NoStop}%
\bibitem [{\citenamefont {Arutyunov}\ and\ \citenamefont {van Tongeren}(2015)}]{Arutyunov:2014jfa}%
  \BibitemOpen
  \bibfield  {author} {\bibinfo {author} {\bibfnamefont {G.}~\bibnamefont {Arutyunov}}\ and\ \bibinfo {author} {\bibfnamefont {S.~J.}\ \bibnamefont {van Tongeren}},\ }\bibfield  {title} {\bibinfo {title} {{Double Wick rotating Green-Schwarz strings}},\ }\href {https://doi.org/10.1007/JHEP05(2015)027} {\bibfield  {journal} {\bibinfo  {journal} {JHEP}\ }\textbf {\bibinfo {volume} {05}},\ \bibinfo {pages} {027}},\ \Eprint {https://arxiv.org/abs/1412.5137} {arXiv:1412.5137 [hep-th]} \BibitemShut {NoStop}%
\bibitem [{\citenamefont {Jones}\ \emph {et~al.}(2004)\citenamefont {Jones}, \citenamefont {Maloney},\ and\ \citenamefont {Strominger}}]{Jones:2004rg}%
  \BibitemOpen
  \bibfield  {author} {\bibinfo {author} {\bibfnamefont {G.}~\bibnamefont {Jones}}, \bibinfo {author} {\bibfnamefont {A.}~\bibnamefont {Maloney}},\ and\ \bibinfo {author} {\bibfnamefont {A.}~\bibnamefont {Strominger}},\ }\bibfield  {title} {\bibinfo {title} {{Nonsingular solutions for S-branes}},\ }\href {https://doi.org/10.1103/PhysRevD.69.126008} {\bibfield  {journal} {\bibinfo  {journal} {Phys. Rev. D}\ }\textbf {\bibinfo {volume} {69}},\ \bibinfo {pages} {126008} (\bibinfo {year} {2004})},\ \Eprint {https://arxiv.org/abs/hep-th/0403050} {arXiv:hep-th/0403050} \BibitemShut {NoStop}%
\bibitem [{\citenamefont {Astorino}(2026)}]{Astorino:2026kuv}%
  \BibitemOpen
  \bibfield  {author} {\bibinfo {author} {\bibfnamefont {M.}~\bibnamefont {Astorino}},\ }\bibfield  {title} {\bibinfo {title} {{Black holes in rotating, electromagnetic backgrounds and topological Kerr-Newman-NUT spacetimes}},\ }\href@noop {} {\  (\bibinfo {year} {2026})},\ \Eprint {https://arxiv.org/abs/2604.05017} {arXiv:2604.05017 [gr-qc]} \BibitemShut {NoStop}%
\bibitem [{\citenamefont {Dai}\ \emph {et~al.}(2025)\citenamefont {Dai}, \citenamefont {Fang},\ and\ \citenamefont {Fujita}}]{Dai:2025yhi}%
  \BibitemOpen
  \bibfield  {author} {\bibinfo {author} {\bibfnamefont {H.-Y.}\ \bibnamefont {Dai}}, \bibinfo {author} {\bibfnamefont {X.-H.}\ \bibnamefont {Fang}},\ and\ \bibinfo {author} {\bibfnamefont {M.}~\bibnamefont {Fujita}},\ }\bibfield  {title} {\bibinfo {title} {{Notes on the double Wick rotated BTZ black hole}},\ }\href {https://doi.org/10.1103/67wf-hqlm} {\bibfield  {journal} {\bibinfo  {journal} {Phys. Rev. D}\ }\textbf {\bibinfo {volume} {112}},\ \bibinfo {pages} {026034} (\bibinfo {year} {2025})},\ \Eprint {https://arxiv.org/abs/2504.10562} {arXiv:2504.10562 [hep-th]} \BibitemShut {NoStop}%
\bibitem [{\citenamefont {Coll{\'e}aux}\ \emph {et~al.}(2025)\citenamefont {Coll{\'e}aux}, \citenamefont {Kol{\'a}{\v{r}}},\ and\ \citenamefont {M{\'a}lek}}]{Colleaux:2025uiw}%
  \BibitemOpen
  \bibfield  {author} {\bibinfo {author} {\bibfnamefont {A.}~\bibnamefont {Coll{\'e}aux}}, \bibinfo {author} {\bibfnamefont {I.}~\bibnamefont {Kol{\'a}{\v{r}}}},\ and\ \bibinfo {author} {\bibfnamefont {T.}~\bibnamefont {M{\'a}lek}},\ }\bibfield  {title} {\bibinfo {title} {{Double Wick rotations between symmetries of Taub-NUT, near-horizon extreme Kerr, and swirling spacetimes}},\ }\href {https://doi.org/10.1103/y6sl-b6x4} {\bibfield  {journal} {\bibinfo  {journal} {Phys. Rev. D}\ }\textbf {\bibinfo {volume} {112}},\ \bibinfo {pages} {124040} (\bibinfo {year} {2025})},\ \Eprint {https://arxiv.org/abs/2509.22309} {arXiv:2509.22309 [gr-qc]} \BibitemShut {NoStop}%
\bibitem [{\citenamefont {Kostelecky}\ and\ \citenamefont {Russell}(2011)}]{Kostelecky:2008ts}%
  \BibitemOpen
  \bibfield  {author} {\bibinfo {author} {\bibfnamefont {V.~A.}\ \bibnamefont {Kostelecky}}\ and\ \bibinfo {author} {\bibfnamefont {N.}~\bibnamefont {Russell}},\ }\bibfield  {title} {\bibinfo {title} {{Data Tables for Lorentz and CPT Violation}},\ }\href {https://doi.org/10.1103/RevModPhys.83.11} {\bibfield  {journal} {\bibinfo  {journal} {Rev. Mod. Phys.}\ }\textbf {\bibinfo {volume} {83}},\ \bibinfo {pages} {11} (\bibinfo {year} {2011})},\ \Eprint {https://arxiv.org/abs/0801.0287} {arXiv:0801.0287 [hep-ph]} \BibitemShut {NoStop}%
\bibitem [{\citenamefont {Mattingly}(2005)}]{Mattingly:2005re}%
  \BibitemOpen
  \bibfield  {author} {\bibinfo {author} {\bibfnamefont {D.}~\bibnamefont {Mattingly}},\ }\bibfield  {title} {\bibinfo {title} {{Modern tests of Lorentz invariance}},\ }\href {https://doi.org/10.12942/lrr-2005-5} {\bibfield  {journal} {\bibinfo  {journal} {Living Rev. Rel.}\ }\textbf {\bibinfo {volume} {8}},\ \bibinfo {pages} {5} (\bibinfo {year} {2005})},\ \Eprint {https://arxiv.org/abs/gr-qc/0502097} {arXiv:gr-qc/0502097} \BibitemShut {NoStop}%
\bibitem [{\citenamefont {Halim}\ \emph {et~al.}(2026)\citenamefont {Halim} \emph {et~al.}}]{PierreAuger:2026dip}%
  \BibitemOpen
  \bibfield  {author} {\bibinfo {author} {\bibfnamefont {A.~A.}\ \bibnamefont {Halim}} \emph {et~al.} (\bibinfo {collaboration} {Pierre Auger}),\ }\bibfield  {title} {\bibinfo {title} {{Bounds on Lorentz invariance violation from muon fluctuations at the Pierre Auger Observatory}},\ }\href@noop {} {\  (\bibinfo {year} {2026})},\ \Eprint {https://arxiv.org/abs/2602.14720} {arXiv:2602.14720 [astro-ph.HE]} \BibitemShut {NoStop}%
\bibitem [{\citenamefont {Kostelecky}(2004)}]{Kostelecky:2003fs}%
  \BibitemOpen
  \bibfield  {author} {\bibinfo {author} {\bibfnamefont {V.~A.}\ \bibnamefont {Kostelecky}},\ }\bibfield  {title} {\bibinfo {title} {{Gravity, Lorentz violation, and the standard model}},\ }\href {https://doi.org/10.1103/PhysRevD.69.105009} {\bibfield  {journal} {\bibinfo  {journal} {Phys. Rev. D}\ }\textbf {\bibinfo {volume} {69}},\ \bibinfo {pages} {105009} (\bibinfo {year} {2004})},\ \Eprint {https://arxiv.org/abs/hep-th/0312310} {arXiv:hep-th/0312310} \BibitemShut {NoStop}%
\bibitem [{\citenamefont {Colladay}\ and\ \citenamefont {Kostelecky}(1998)}]{Colladay:1998fq}%
  \BibitemOpen
  \bibfield  {author} {\bibinfo {author} {\bibfnamefont {D.}~\bibnamefont {Colladay}}\ and\ \bibinfo {author} {\bibfnamefont {V.~A.}\ \bibnamefont {Kostelecky}},\ }\bibfield  {title} {\bibinfo {title} {{Lorentz violating extension of the standard model}},\ }\href {https://doi.org/10.1103/PhysRevD.58.116002} {\bibfield  {journal} {\bibinfo  {journal} {Phys. Rev. D}\ }\textbf {\bibinfo {volume} {58}},\ \bibinfo {pages} {116002} (\bibinfo {year} {1998})},\ \Eprint {https://arxiv.org/abs/hep-ph/9809521} {arXiv:hep-ph/9809521} \BibitemShut {NoStop}%
\bibitem [{\citenamefont {Seifert}(2019)}]{Seifert:2018mmr}%
  \BibitemOpen
  \bibfield  {author} {\bibinfo {author} {\bibfnamefont {M.~D.}\ \bibnamefont {Seifert}},\ }\bibfield  {title} {\bibinfo {title} {{Constraints and degrees of freedom in Lorentz-violating field theories}},\ }\href {https://doi.org/10.1103/PhysRevD.99.045003} {\bibfield  {journal} {\bibinfo  {journal} {Phys. Rev. D}\ }\textbf {\bibinfo {volume} {99}},\ \bibinfo {pages} {045003} (\bibinfo {year} {2019})},\ \Eprint {https://arxiv.org/abs/1810.09512} {arXiv:1810.09512 [hep-th]} \BibitemShut {NoStop}%
\bibitem [{\citenamefont {J{\'u}nior}\ \emph {et~al.}(2026)\citenamefont {J{\'u}nior}, \citenamefont {Felipe}, \citenamefont {Ba{\^e}ta~Scarpelli}, \citenamefont {Petrov},\ and\ \citenamefont {Helay{\"e}l-Neto}}]{Junior:2026gta}%
  \BibitemOpen
  \bibfield  {author} {\bibinfo {author} {\bibfnamefont {E.~N.}\ \bibnamefont {J{\'u}nior}}, \bibinfo {author} {\bibfnamefont {J.~C.~C.}\ \bibnamefont {Felipe}}, \bibinfo {author} {\bibfnamefont {A.~P.}\ \bibnamefont {Ba{\^e}ta~Scarpelli}}, \bibinfo {author} {\bibfnamefont {A.~Y.}\ \bibnamefont {Petrov}},\ and\ \bibinfo {author} {\bibfnamefont {J.~A.}\ \bibnamefont {Helay{\"e}l-Neto}},\ }\bibfield  {title} {\bibinfo {title} {{Classical investigations in a CPT-even Lorentz-violating model and their implications for the Compton effect}},\ }\href@noop {} {\  (\bibinfo {year} {2026})},\ \Eprint {https://arxiv.org/abs/2602.23456} {arXiv:2602.23456 [hep-th]} \BibitemShut {NoStop}%
\bibitem [{\citenamefont {Alfaia}\ \emph {et~al.}(2026)\citenamefont {Alfaia}, \citenamefont {Carvalho}, \citenamefont {Lehum}, \citenamefont {Nascimento}, \citenamefont {Petrov},\ and\ \citenamefont {Porf{\'\i}rio}}]{Alfaia:2026swy}%
  \BibitemOpen
  \bibfield  {author} {\bibinfo {author} {\bibfnamefont {R.~B.}\ \bibnamefont {Alfaia}}, \bibinfo {author} {\bibfnamefont {W.}~\bibnamefont {Carvalho}}, \bibinfo {author} {\bibfnamefont {A.~C.}\ \bibnamefont {Lehum}}, \bibinfo {author} {\bibfnamefont {J.~R.}\ \bibnamefont {Nascimento}}, \bibinfo {author} {\bibfnamefont {A.~Y.}\ \bibnamefont {Petrov}},\ and\ \bibinfo {author} {\bibfnamefont {P.~J.}\ \bibnamefont {Porf{\'\i}rio}},\ }\bibfield  {title} {\bibinfo {title} {{Lorentz violating quadratic gravity}},\ }\href@noop {} {\  (\bibinfo {year} {2026})},\ \Eprint {https://arxiv.org/abs/2603.02980} {arXiv:2603.02980 [hep-th]} \BibitemShut {NoStop}%
\bibitem [{\citenamefont {Kosteleck{\'y}}\ and\ \citenamefont {Mewes}(2018)}]{Kostelecky:2017zob}%
  \BibitemOpen
  \bibfield  {author} {\bibinfo {author} {\bibfnamefont {V.~A.}\ \bibnamefont {Kosteleck{\'y}}}\ and\ \bibinfo {author} {\bibfnamefont {M.}~\bibnamefont {Mewes}},\ }\bibfield  {title} {\bibinfo {title} {{Lorentz and Diffeomorphism Violations in Linearized Gravity}},\ }\href {https://doi.org/10.1016/j.physletb.2018.01.082} {\bibfield  {journal} {\bibinfo  {journal} {Phys. Lett. B}\ }\textbf {\bibinfo {volume} {779}},\ \bibinfo {pages} {136} (\bibinfo {year} {2018})},\ \Eprint {https://arxiv.org/abs/1712.10268} {arXiv:1712.10268 [gr-qc]} \BibitemShut {NoStop}%
\bibitem [{\citenamefont {Bellorin}\ \emph {et~al.}(2013)\citenamefont {Bellorin}, \citenamefont {Restuccia},\ and\ \citenamefont {Sotomayor}}]{Bellorin:2013zbp}%
  \BibitemOpen
  \bibfield  {author} {\bibinfo {author} {\bibfnamefont {J.}~\bibnamefont {Bellorin}}, \bibinfo {author} {\bibfnamefont {A.}~\bibnamefont {Restuccia}},\ and\ \bibinfo {author} {\bibfnamefont {A.}~\bibnamefont {Sotomayor}},\ }\bibfield  {title} {\bibinfo {title} {{Consistent Ho{\v{r}}ava gravity without extra modes and equivalent to general relativity at the linearized level}},\ }\href {https://doi.org/10.1103/PhysRevD.87.084020} {\bibfield  {journal} {\bibinfo  {journal} {Phys. Rev. D}\ }\textbf {\bibinfo {volume} {87}},\ \bibinfo {pages} {084020} (\bibinfo {year} {2013})},\ \Eprint {https://arxiv.org/abs/1302.1357} {arXiv:1302.1357 [hep-th]} \BibitemShut {NoStop}%
\bibitem [{\citenamefont {Horava}(2009{\natexlab{a}})}]{Horava:2009if}%
  \BibitemOpen
  \bibfield  {author} {\bibinfo {author} {\bibfnamefont {P.}~\bibnamefont {Horava}},\ }\bibfield  {title} {\bibinfo {title} {{Spectral Dimension of the Universe in Quantum Gravity at a Lifshitz Point}},\ }\href {https://doi.org/10.1103/PhysRevLett.102.161301} {\bibfield  {journal} {\bibinfo  {journal} {Phys. Rev. Lett.}\ }\textbf {\bibinfo {volume} {102}},\ \bibinfo {pages} {161301} (\bibinfo {year} {2009}{\natexlab{a}})},\ \Eprint {https://arxiv.org/abs/0902.3657} {arXiv:0902.3657 [hep-th]} \BibitemShut {NoStop}%
\bibitem [{\citenamefont {Horava}(2009{\natexlab{b}})}]{Horava:2009uw}%
  \BibitemOpen
  \bibfield  {author} {\bibinfo {author} {\bibfnamefont {P.}~\bibnamefont {Horava}},\ }\bibfield  {title} {\bibinfo {title} {{Quantum Gravity at a Lifshitz Point}},\ }\href {https://doi.org/10.1103/PhysRevD.79.084008} {\bibfield  {journal} {\bibinfo  {journal} {Phys. Rev. D}\ }\textbf {\bibinfo {volume} {79}},\ \bibinfo {pages} {084008} (\bibinfo {year} {2009}{\natexlab{b}})},\ \Eprint {https://arxiv.org/abs/0901.3775} {arXiv:0901.3775 [hep-th]} \BibitemShut {NoStop}%
\bibitem [{\citenamefont {Barvinsky}\ \emph {et~al.}(2016)\citenamefont {Barvinsky}, \citenamefont {Blas}, \citenamefont {Herrero-Valea}, \citenamefont {Sibiryakov},\ and\ \citenamefont {Steinwachs}}]{Barvinsky:2015kil}%
  \BibitemOpen
  \bibfield  {author} {\bibinfo {author} {\bibfnamefont {A.~O.}\ \bibnamefont {Barvinsky}}, \bibinfo {author} {\bibfnamefont {D.}~\bibnamefont {Blas}}, \bibinfo {author} {\bibfnamefont {M.}~\bibnamefont {Herrero-Valea}}, \bibinfo {author} {\bibfnamefont {S.~M.}\ \bibnamefont {Sibiryakov}},\ and\ \bibinfo {author} {\bibfnamefont {C.~F.}\ \bibnamefont {Steinwachs}},\ }\bibfield  {title} {\bibinfo {title} {{Renormalization of Ho{\v{r}}ava gravity}},\ }\href {https://doi.org/10.1103/PhysRevD.93.064022} {\bibfield  {journal} {\bibinfo  {journal} {Phys. Rev. D}\ }\textbf {\bibinfo {volume} {93}},\ \bibinfo {pages} {064022} (\bibinfo {year} {2016})},\ \Eprint {https://arxiv.org/abs/1512.02250} {arXiv:1512.02250 [hep-th]} \BibitemShut {NoStop}%
\bibitem [{\citenamefont {Anselmi}\ and\ \citenamefont {Halat}(2007)}]{Anselmi:2007ri}%
  \BibitemOpen
  \bibfield  {author} {\bibinfo {author} {\bibfnamefont {D.}~\bibnamefont {Anselmi}}\ and\ \bibinfo {author} {\bibfnamefont {M.}~\bibnamefont {Halat}},\ }\bibfield  {title} {\bibinfo {title} {{Renormalization of Lorentz violating theories}},\ }\href {https://doi.org/10.1103/PhysRevD.76.125011} {\bibfield  {journal} {\bibinfo  {journal} {Phys. Rev. D}\ }\textbf {\bibinfo {volume} {76}},\ \bibinfo {pages} {125011} (\bibinfo {year} {2007})},\ \Eprint {https://arxiv.org/abs/0707.2480} {arXiv:0707.2480 [hep-th]} \BibitemShut {NoStop}%
\bibitem [{\citenamefont {Blas}\ \emph {et~al.}(2010)\citenamefont {Blas}, \citenamefont {Pujolas},\ and\ \citenamefont {Sibiryakov}}]{Blas:2009qj}%
  \BibitemOpen
  \bibfield  {author} {\bibinfo {author} {\bibfnamefont {D.}~\bibnamefont {Blas}}, \bibinfo {author} {\bibfnamefont {O.}~\bibnamefont {Pujolas}},\ and\ \bibinfo {author} {\bibfnamefont {S.}~\bibnamefont {Sibiryakov}},\ }\bibfield  {title} {\bibinfo {title} {{Consistent Extension of Horava Gravity}},\ }\href {https://doi.org/10.1103/PhysRevLett.104.181302} {\bibfield  {journal} {\bibinfo  {journal} {Phys. Rev. Lett.}\ }\textbf {\bibinfo {volume} {104}},\ \bibinfo {pages} {181302} (\bibinfo {year} {2010})},\ \Eprint {https://arxiv.org/abs/0909.3525} {arXiv:0909.3525 [hep-th]} \BibitemShut {NoStop}%
\bibitem [{\citenamefont {Blas}\ \emph {et~al.}(2011)\citenamefont {Blas}, \citenamefont {Pujolas},\ and\ \citenamefont {Sibiryakov}}]{Blas:2010hb}%
  \BibitemOpen
  \bibfield  {author} {\bibinfo {author} {\bibfnamefont {D.}~\bibnamefont {Blas}}, \bibinfo {author} {\bibfnamefont {O.}~\bibnamefont {Pujolas}},\ and\ \bibinfo {author} {\bibfnamefont {S.}~\bibnamefont {Sibiryakov}},\ }\bibfield  {title} {\bibinfo {title} {{Models of non-relativistic quantum gravity: The Good, the bad and the healthy}},\ }\href {https://doi.org/10.1007/JHEP04(2011)018} {\bibfield  {journal} {\bibinfo  {journal} {JHEP}\ }\textbf {\bibinfo {volume} {04}},\ \bibinfo {pages} {018}},\ \Eprint {https://arxiv.org/abs/1007.3503} {arXiv:1007.3503 [hep-th]} \BibitemShut {NoStop}%
\bibitem [{\citenamefont {Calcagni}(2009)}]{Calcagni:2009ar}%
  \BibitemOpen
  \bibfield  {author} {\bibinfo {author} {\bibfnamefont {G.}~\bibnamefont {Calcagni}},\ }\bibfield  {title} {\bibinfo {title} {{Cosmology of the Lifshitz universe}},\ }\href {https://doi.org/10.1088/1126-6708/2009/09/112} {\bibfield  {journal} {\bibinfo  {journal} {JHEP}\ }\textbf {\bibinfo {volume} {09}},\ \bibinfo {pages} {112}},\ \Eprint {https://arxiv.org/abs/0904.0829} {arXiv:0904.0829 [hep-th]} \BibitemShut {NoStop}%
\bibitem [{\citenamefont {Kiritsis}\ and\ \citenamefont {Kofinas}(2009)}]{Kiritsis:2009sh}%
  \BibitemOpen
  \bibfield  {author} {\bibinfo {author} {\bibfnamefont {E.}~\bibnamefont {Kiritsis}}\ and\ \bibinfo {author} {\bibfnamefont {G.}~\bibnamefont {Kofinas}},\ }\bibfield  {title} {\bibinfo {title} {{Horava-Lifshitz Cosmology}},\ }\href {https://doi.org/10.1016/j.nuclphysb.2009.05.005} {\bibfield  {journal} {\bibinfo  {journal} {Nucl. Phys. B}\ }\textbf {\bibinfo {volume} {821}},\ \bibinfo {pages} {467} (\bibinfo {year} {2009})},\ \Eprint {https://arxiv.org/abs/0904.1334} {arXiv:0904.1334 [hep-th]} \BibitemShut {NoStop}%
\bibitem [{\citenamefont {Vilenkin}(1982)}]{Vilenkin:1982de}%
  \BibitemOpen
  \bibfield  {author} {\bibinfo {author} {\bibfnamefont {A.}~\bibnamefont {Vilenkin}},\ }\bibfield  {title} {\bibinfo {title} {{Creation of Universes from Nothing}},\ }\href {https://doi.org/10.1016/0370-2693(82)90866-8} {\bibfield  {journal} {\bibinfo  {journal} {Phys. Lett. B}\ }\textbf {\bibinfo {volume} {117}},\ \bibinfo {pages} {25} (\bibinfo {year} {1982})}\BibitemShut {NoStop}%
\bibitem [{\citenamefont {Pereira}\ \emph {et~al.}(2007)\citenamefont {Pereira}, \citenamefont {Pitrou},\ and\ \citenamefont {Uzan}}]{Pereira:2007yy}%
  \BibitemOpen
  \bibfield  {author} {\bibinfo {author} {\bibfnamefont {T.~S.}\ \bibnamefont {Pereira}}, \bibinfo {author} {\bibfnamefont {C.}~\bibnamefont {Pitrou}},\ and\ \bibinfo {author} {\bibfnamefont {J.-P.}\ \bibnamefont {Uzan}},\ }\bibfield  {title} {\bibinfo {title} {{Theory of cosmological perturbations in an anisotropic universe}},\ }\href {https://doi.org/10.1088/1475-7516/2007/09/006} {\bibfield  {journal} {\bibinfo  {journal} {JCAP}\ }\textbf {\bibinfo {volume} {09}},\ \bibinfo {pages} {006}},\ \Eprint {https://arxiv.org/abs/0707.0736} {arXiv:0707.0736 [astro-ph]} \BibitemShut {NoStop}%
\bibitem [{\citenamefont {Mukohyama}(2010)}]{Mukohyama:2010xz}%
  \BibitemOpen
  \bibfield  {author} {\bibinfo {author} {\bibfnamefont {S.}~\bibnamefont {Mukohyama}},\ }\bibfield  {title} {\bibinfo {title} {{Horava-Lifshitz Cosmology: A Review}},\ }\href {https://doi.org/10.1088/0264-9381/27/22/223101} {\bibfield  {journal} {\bibinfo  {journal} {Class. Quant. Grav.}\ }\textbf {\bibinfo {volume} {27}},\ \bibinfo {pages} {223101} (\bibinfo {year} {2010})},\ \Eprint {https://arxiv.org/abs/1007.5199} {arXiv:1007.5199 [hep-th]} \BibitemShut {NoStop}%
\bibitem [{\citenamefont {Golovnev}\ \emph {et~al.}(2008)\citenamefont {Golovnev}, \citenamefont {Mukhanov},\ and\ \citenamefont {Vanchurin}}]{Golovnev:2008hv}%
  \BibitemOpen
  \bibfield  {author} {\bibinfo {author} {\bibfnamefont {A.}~\bibnamefont {Golovnev}}, \bibinfo {author} {\bibfnamefont {V.}~\bibnamefont {Mukhanov}},\ and\ \bibinfo {author} {\bibfnamefont {V.}~\bibnamefont {Vanchurin}},\ }\bibfield  {title} {\bibinfo {title} {{Gravitational waves in vector inflation}},\ }\href {https://doi.org/10.1088/1475-7516/2008/11/018} {\bibfield  {journal} {\bibinfo  {journal} {JCAP}\ }\textbf {\bibinfo {volume} {11}},\ \bibinfo {pages} {018}},\ \Eprint {https://arxiv.org/abs/0810.4304} {arXiv:0810.4304 [astro-ph]} \BibitemShut {NoStop}%
\bibitem [{\citenamefont {Kobayashi}\ and\ \citenamefont {Yokoyama}(2009)}]{Kobayashi:2009hj}%
  \BibitemOpen
  \bibfield  {author} {\bibinfo {author} {\bibfnamefont {T.}~\bibnamefont {Kobayashi}}\ and\ \bibinfo {author} {\bibfnamefont {S.}~\bibnamefont {Yokoyama}},\ }\bibfield  {title} {\bibinfo {title} {{Gravitational waves from p-form inflation}},\ }\href {https://doi.org/10.1088/1475-7516/2009/05/004} {\bibfield  {journal} {\bibinfo  {journal} {JCAP}\ }\textbf {\bibinfo {volume} {05}},\ \bibinfo {pages} {004}},\ \Eprint {https://arxiv.org/abs/0903.2769} {arXiv:0903.2769 [astro-ph.CO]} \BibitemShut {NoStop}%
\bibitem [{\citenamefont {De~Felice}\ and\ \citenamefont {Hell}(2025{\natexlab{a}})}]{DeFelice:2025ykh}%
  \BibitemOpen
  \bibfield  {author} {\bibinfo {author} {\bibfnamefont {A.}~\bibnamefont {De~Felice}}\ and\ \bibinfo {author} {\bibfnamefont {A.}~\bibnamefont {Hell}},\ }\bibfield  {title} {\bibinfo {title} {{On the cosmological degrees of freedom of Proca field with non-minimal coupling to gravity}},\ }\href {https://doi.org/10.1007/JHEP07(2025)228} {\bibfield  {journal} {\bibinfo  {journal} {JHEP}\ }\textbf {\bibinfo {volume} {07}},\ \bibinfo {pages} {228}},\ \Eprint {https://arxiv.org/abs/2503.07454} {arXiv:2503.07454 [gr-qc]} \BibitemShut {NoStop}%
\bibitem [{\citenamefont {De~Felice}\ and\ \citenamefont {Hell}(2025{\natexlab{b}})}]{DeFelice:2025khe}%
  \BibitemOpen
  \bibfield  {author} {\bibinfo {author} {\bibfnamefont {A.}~\bibnamefont {De~Felice}}\ and\ \bibinfo {author} {\bibfnamefont {A.}~\bibnamefont {Hell}},\ }\bibfield  {title} {\bibinfo {title} {{The non-minimal 3-form cosmology and the rise of the cuscuton}},\ }\href {https://doi.org/10.1007/JHEP11(2025)132} {\bibfield  {journal} {\bibinfo  {journal} {JHEP}\ }\textbf {\bibinfo {volume} {11}},\ \bibinfo {pages} {132}},\ \Eprint {https://arxiv.org/abs/2509.02323} {arXiv:2509.02323 [gr-qc]} \BibitemShut {NoStop}%
\bibitem [{\citenamefont {Hell}\ and\ \citenamefont {Daniel}(2026)}]{Hell:2026sxt}%
  \BibitemOpen
  \bibfield  {author} {\bibinfo {author} {\bibfnamefont {A.}~\bibnamefont {Hell}}\ and\ \bibinfo {author} {\bibfnamefont {T.}~\bibnamefont {Daniel}},\ }\bibfield  {title} {\bibinfo {title} {{Branching Universes}},\ }\href@noop {} {\  (\bibinfo {year} {2026})},\ \Eprint {https://arxiv.org/abs/2603.18147} {arXiv:2603.18147 [hep-th]} \BibitemShut {NoStop}%
\bibitem [{\citenamefont {Lin}\ and\ \citenamefont {Mukohyama}(2017)}]{Lin:2017oow}%
  \BibitemOpen
  \bibfield  {author} {\bibinfo {author} {\bibfnamefont {C.}~\bibnamefont {Lin}}\ and\ \bibinfo {author} {\bibfnamefont {S.}~\bibnamefont {Mukohyama}},\ }\bibfield  {title} {\bibinfo {title} {{A Class of Minimally Modified Gravity Theories}},\ }\href {https://doi.org/10.1088/1475-7516/2017/10/033} {\bibfield  {journal} {\bibinfo  {journal} {JCAP}\ }\textbf {\bibinfo {volume} {10}},\ \bibinfo {pages} {033}},\ \Eprint {https://arxiv.org/abs/1708.03757} {arXiv:1708.03757 [gr-qc]} \BibitemShut {NoStop}%
\bibitem [{\citenamefont {Ostrogradsky}(1850)}]{Ostrogradsky:1850fid}%
  \BibitemOpen
  \bibfield  {author} {\bibinfo {author} {\bibfnamefont {M.}~\bibnamefont {Ostrogradsky}},\ }\bibfield  {title} {\bibinfo {title} {{M{\'e}moires sur les {\'e}quations diff{\'e}rentielles, relatives au probl{\`e}me des isop{\'e}rim{\`e}tres}},\ }\href@noop {} {\bibfield  {journal} {\bibinfo  {journal} {Mem. Acad. St. Petersbourg}\ }\textbf {\bibinfo {volume} {6}},\ \bibinfo {pages} {385} (\bibinfo {year} {1850})}\BibitemShut {NoStop}%
\bibitem [{\citenamefont {Dvali}\ \emph {et~al.}(2005)\citenamefont {Dvali}, \citenamefont {Papucci},\ and\ \citenamefont {Schwartz}}]{Dvali:2005nt}%
  \BibitemOpen
  \bibfield  {author} {\bibinfo {author} {\bibfnamefont {G.}~\bibnamefont {Dvali}}, \bibinfo {author} {\bibfnamefont {M.}~\bibnamefont {Papucci}},\ and\ \bibinfo {author} {\bibfnamefont {M.~D.}\ \bibnamefont {Schwartz}},\ }\bibfield  {title} {\bibinfo {title} {{Infrared Lorentz violation and slowly instantaneous electricity}},\ }\href {https://doi.org/10.1103/PhysRevLett.94.191602} {\bibfield  {journal} {\bibinfo  {journal} {Phys. Rev. Lett.}\ }\textbf {\bibinfo {volume} {94}},\ \bibinfo {pages} {191602} (\bibinfo {year} {2005})},\ \Eprint {https://arxiv.org/abs/hep-th/0501157} {arXiv:hep-th/0501157} \BibitemShut {NoStop}%
\bibitem [{\citenamefont {Hell}\ \emph {et~al.}(2026)\citenamefont {Hell}, \citenamefont {Ferreira}, \citenamefont {L{\"u}st},\ and\ \citenamefont {Sasaki}}]{Hell:2026blj}%
  \BibitemOpen
  \bibfield  {author} {\bibinfo {author} {\bibfnamefont {A.}~\bibnamefont {Hell}}, \bibinfo {author} {\bibfnamefont {E.~G.~M.}\ \bibnamefont {Ferreira}}, \bibinfo {author} {\bibfnamefont {D.}~\bibnamefont {L{\"u}st}},\ and\ \bibinfo {author} {\bibfnamefont {M.}~\bibnamefont {Sasaki}},\ }\bibfield  {title} {\bibinfo {title} {{The recipe for the degrees of freedom}},\ }\href {https://doi.org/10.1007/JHEP03(2026)235} {\bibfield  {journal} {\bibinfo  {journal} {JHEP}\ }\textbf {\bibinfo {volume} {03}},\ \bibinfo {pages} {235}},\ \Eprint {https://arxiv.org/abs/2601.10288} {arXiv:2601.10288 [hep-th]} \BibitemShut {NoStop}%
\bibitem [{\citenamefont {Vainshtein}(1972)}]{Vainshtein:1972sx}%
  \BibitemOpen
  \bibfield  {author} {\bibinfo {author} {\bibfnamefont {A.~I.}\ \bibnamefont {Vainshtein}},\ }\bibfield  {title} {\bibinfo {title} {{To the problem of nonvanishing gravitation mass}},\ }\href {https://doi.org/10.1016/0370-2693(72)90147-5} {\bibfield  {journal} {\bibinfo  {journal} {Phys. Lett. B}\ }\textbf {\bibinfo {volume} {39}},\ \bibinfo {pages} {393} (\bibinfo {year} {1972})}\BibitemShut {NoStop}%
\bibitem [{\citenamefont {Gruzinov}(2005)}]{Gruzinov:2001hp}%
  \BibitemOpen
  \bibfield  {author} {\bibinfo {author} {\bibfnamefont {A.}~\bibnamefont {Gruzinov}},\ }\bibfield  {title} {\bibinfo {title} {{On the graviton mass}},\ }\href {https://doi.org/10.1016/j.newast.2004.12.001} {\bibfield  {journal} {\bibinfo  {journal} {New Astron.}\ }\textbf {\bibinfo {volume} {10}},\ \bibinfo {pages} {311} (\bibinfo {year} {2005})},\ \Eprint {https://arxiv.org/abs/astro-ph/0112246} {arXiv:astro-ph/0112246} \BibitemShut {NoStop}%
\bibitem [{\citenamefont {Deffayet}\ \emph {et~al.}(2002)\citenamefont {Deffayet}, \citenamefont {Dvali}, \citenamefont {Gabadadze},\ and\ \citenamefont {Vainshtein}}]{Deffayet:2001uk}%
  \BibitemOpen
  \bibfield  {author} {\bibinfo {author} {\bibfnamefont {C.}~\bibnamefont {Deffayet}}, \bibinfo {author} {\bibfnamefont {G.~R.}\ \bibnamefont {Dvali}}, \bibinfo {author} {\bibfnamefont {G.}~\bibnamefont {Gabadadze}},\ and\ \bibinfo {author} {\bibfnamefont {A.~I.}\ \bibnamefont {Vainshtein}},\ }\bibfield  {title} {\bibinfo {title} {{Nonperturbative continuity in graviton mass versus perturbative discontinuity}},\ }\href {https://doi.org/10.1103/PhysRevD.65.044026} {\bibfield  {journal} {\bibinfo  {journal} {Phys. Rev. D}\ }\textbf {\bibinfo {volume} {65}},\ \bibinfo {pages} {044026} (\bibinfo {year} {2002})},\ \Eprint {https://arxiv.org/abs/hep-th/0106001} {arXiv:hep-th/0106001} \BibitemShut {NoStop}%
\bibitem [{\citenamefont {Chamseddine}\ and\ \citenamefont {Mukhanov}(2018)}]{Chamseddine:2018gqh}%
  \BibitemOpen
  \bibfield  {author} {\bibinfo {author} {\bibfnamefont {A.~H.}\ \bibnamefont {Chamseddine}}\ and\ \bibinfo {author} {\bibfnamefont {V.}~\bibnamefont {Mukhanov}},\ }\bibfield  {title} {\bibinfo {title} {{Mimetic Massive Gravity: Beyond Linear Approximation}},\ }\href {https://doi.org/10.1007/JHEP06(2018)062} {\bibfield  {journal} {\bibinfo  {journal} {JHEP}\ }\textbf {\bibinfo {volume} {06}},\ \bibinfo {pages} {062}},\ \Eprint {https://arxiv.org/abs/1805.06598} {arXiv:1805.06598 [hep-th]} \BibitemShut {NoStop}%
\bibitem [{\citenamefont {Hell}(2022{\natexlab{a}})}]{Hell:2021wzm}%
  \BibitemOpen
  \bibfield  {author} {\bibinfo {author} {\bibfnamefont {A.}~\bibnamefont {Hell}},\ }\bibfield  {title} {\bibinfo {title} {{On the duality of massive Kalb-Ramond and Proca fields}},\ }\href {https://doi.org/10.1088/1475-7516/2022/01/056} {\bibfield  {journal} {\bibinfo  {journal} {JCAP}\ }\textbf {\bibinfo {volume} {01}}\bibfield  {number} {\bibinfo  {number} { (01)},\ \bibinfo {pages} {056}},\ }\Eprint {https://arxiv.org/abs/2109.05030} {arXiv:2109.05030 [hep-th]} \BibitemShut {NoStop}%
\bibitem [{\citenamefont {Hell}(2022{\natexlab{b}})}]{Hell:2021oea}%
  \BibitemOpen
  \bibfield  {author} {\bibinfo {author} {\bibfnamefont {A.}~\bibnamefont {Hell}},\ }\bibfield  {title} {\bibinfo {title} {{The strong couplings of massive Yang-Mills theory}},\ }\href {https://doi.org/10.1007/JHEP03(2022)167} {\bibfield  {journal} {\bibinfo  {journal} {JHEP}\ }\textbf {\bibinfo {volume} {03}},\ \bibinfo {pages} {167}},\ \Eprint {https://arxiv.org/abs/2111.00017} {arXiv:2111.00017 [hep-th]} \BibitemShut {NoStop}%
\bibitem [{\citenamefont {Hell}(2025)}]{Hell:2025pso}%
  \BibitemOpen
  \bibfield  {author} {\bibinfo {author} {\bibfnamefont {A.}~\bibnamefont {Hell}},\ }\href {https://doi.org/10.1007/978-3-032-01090-2} {\emph {\bibinfo {title} {{The Massless Limit of Massive Gauge Theories}: {To the Strong Coupling and Beyond}}}},\ Springer Theses\ (\bibinfo  {publisher} {Springer Cham},\ \bibinfo {year} {2025})\BibitemShut {NoStop}%
\bibitem [{\citenamefont {Boulware}\ and\ \citenamefont {Gilbert}(1962)}]{PhysRev.126.1563}%
  \BibitemOpen
  \bibfield  {author} {\bibinfo {author} {\bibfnamefont {D.~G.}\ \bibnamefont {Boulware}}\ and\ \bibinfo {author} {\bibfnamefont {W.}~\bibnamefont {Gilbert}},\ }\bibfield  {title} {\bibinfo {title} {Connection between gauge invariance and mass},\ }\href {https://doi.org/10.1103/PhysRev.126.1563} {\bibfield  {journal} {\bibinfo  {journal} {Phys. Rev.}\ }\textbf {\bibinfo {volume} {126}},\ \bibinfo {pages} {1563} (\bibinfo {year} {1962})}\BibitemShut {NoStop}%
\end{thebibliography}%

\end{document}